\documentclass[fleqn,11pt]{article}
\usepackage{amscd,amsmath,amssymb,verbatim}
\usepackage{amsthm}
\usepackage[pdftex]{graphicx}
\usepackage{epstopdf}
\usepackage[T1]{fontenc}
\usepackage{ae,aecompl}
\usepackage{subfigure}
\usepackage{array}
\usepackage{mathrsfs}
\usepackage[nohead]{geometry}
\usepackage[singlespacing]{setspace}
\usepackage{rotating}
\usepackage{multirow}
\usepackage{threeparttable}
\usepackage{enumerate}
\usepackage{subfigure}
\usepackage{placeins}
\usepackage{color, colortbl}
\usepackage{accents}
\usepackage{pdflscape}
\usepackage{natbib}
\usepackage{float}
\numberwithin{equation}{section}

\usepackage{hyperref}

\makeatletter
\newcommand{\distas}[1]{\mathbin{\overset{#1}{\kern\z@\sim}}}%
\newsavebox{\mybox}\newsavebox{\mysim}
\newcommand{\distras}[1]{%
  \savebox{\mybox}{\hbox{\kern3pt$\scriptstyle#1$\kern3pt}}%
  \savebox{\mysim}{\hbox{$\sim$}}%
  \mathbin{\overset{#1}{\kern\z@\resizebox{\wd\mybox}{\ht\mysim}{$\sim$}}}%
}
\makeatother

\renewcommand{\hat}[1]{\widehat{\text{$#1$}}}
\makeatletter
\newsavebox\myboxA
\newsavebox\myboxB
\newlength\mylenA

\renewcommand*\bar[2][0.85]{%
    \sbox{\myboxA}{$\m@th#2$}%
    \setbox\myboxB\null
    \ht\myboxB=\ht\myboxA%
    \dp\myboxB=\dp\myboxA%
    \wd\myboxB=#1\wd\myboxA
    \sbox\myboxB{$\m@th\overline{\copy\myboxB}$}
    \setlength\mylenA{\the\wd\myboxA}
    \addtolength\mylenA{-\the\wd\myboxB}%
    \ifdim\wd\myboxB<\wd\myboxA%
       \rlap{\hskip 0.5\mylenA\usebox\myboxB}{\usebox\myboxA}%
    \else
        \hskip -0.5\mylenA\rlap{\usebox\myboxA}{\hskip 0.5\mylenA\usebox\myboxB}%
    \fi}
\makeatother
\makeatletter
\def\@biblabel#1{\hspace*{-\labelsep}}
\makeatother
\hoffset 
\voffset
\author{Raffaele Mattera\thanks{Sapienza University of Rome, Italy, email: raffaele.mattera@uniroma1.it, ORCiD: 0000-0001-8770-7049} \and Philipp Otto\thanks{Leibniz University, Hannover, Germany, email: philipp.otto@ikg.uni-hannover.de, ORCiD: 0000-0002-9796-6682}}\medskip

\date{\today}
\title{Network log-ARCH models for forecasting stock market volatility}

\begin{document}
\maketitle
\sloppy

\singlespacing

\begin{abstract}
\noindent 
This paper presents a novel dynamic network autoregressive conditional heteroscedasticity (ARCH) model based on spatiotemporal ARCH models to forecast volatility in the US stock market. To improve the forecasting accuracy, the model integrates temporally lagged volatility information and information from adjacent nodes, which may instantaneously spill across the entire network. The model is also suitable for high-dimensional cases where multivariate ARCH models are typically no longer applicable. We adopt the theoretical foundations from spatiotemporal statistics and transfer the dynamic ARCH model for processes to networks. This new approach is compared with independent univariate log-ARCH models. We could quantify the improvements due to the instantaneous network ARCH effects, which are studied for the first time in this paper. The edges are determined based on various distance and correlation measures between the time series. The performances of the alternative networks' definitions are compared in terms of out-of-sample accuracy. Furthermore, we consider ensemble forecasts based on different network definitions.
\end{abstract}

\onehalfspacing


\section{Introduction}

Forecasting volatility in the stock market is crucial, as it provides relevant information for investment and risk management purposes. Volatility forecasting usually employs generalised autoregressive conditional heteroscedasticity (GARCH) models and their extensions \citep{bollerslev1986generalized,andersen1998answering,francq2019garch}. However, the 
use of exponential GARCH models is usually preferred nowadays.  \cite{geweke1986commet} advocates using the log-ARCH model instead of the non-exponential GARCH when dealing with highly persistent volatility processes, large jumps in the data, outliers and skewness.  Moreover, the logarithmic specification used in the log-ARCH model ensures the positivity of the volatility dynamics without imposing further constraints on the parameters and it captures the properties of volatility better than the quadratic specification employed in the standard GARCH. The use of the log-ARCH model has been limited by the problem of dealing with zero values; a potential solution has recently been proposed by \cite{sucarrat2016estimation}. In this paper, we focus on log-ARCH and develop a novel modelling framework, which takes inspiration from spatiotemporal statistics to improve the out-of-sample volatility forecasting performance.

Considering volatility modelling, \cite{otto2018generalised} introduced spatial ARCH (spARCH) models, while \cite{otto2021stochastic} analysed its properties in detail. Further, \cite{sato2017spatial} suggested a logarithmic expression of the volatility equation, which is the equivalent of a log-ARCH model for spatial data. In \cite{sato2021spatial}, the spatial log-ARCH model has been generalised to a spatial log-GARCH model. \cite{otto2022general} introduced a generalization of the spARCH model in a unified framework, allowing for a variety of possible spatial GARCH-type models. More recently, \cite{otto2022dynamic} proposed a dynamic spatiotemporal log-ARCH approach for modelling house prices in Berlin. In this paper, we propose using the \cite{otto2022dynamic} spatiotemporal modelling approach for forecasting stock market volatility. The out-of-sample forecasting performance is compared with the benchmark time-series log-ARCH models and ensemble forecasts.

Although spatial and spatiotemporal modelling is popular in many domains such as epidemiology \citep{sahu2022bayesian,mattera2022weighted}, environmental sciences \citep{huang2011class,cameletti2019bayesian,piter2022helsinki,fasso2022spatiotemporal}, real estate economics \citep{holly2010spatio,baltagi2014further,otto2018spatiotemporal}, it is less explored in finance. By modelling volatility with a spatiotemporal approach, we assume that risk is influenced not only by the temporal fluctuations but also by dependencies with other stocks that are close in the geographical space or, more generally, are in some sense similar.  

According to \cite{pirinsky2006does}, two main explanations of why adjacent information can be helpful for modelling stock market data can be found in local investors' correlated trading activities and by the presence of locally correlated fundamentals. Nevertheless, many studies argue that the geographical distances have quite limited relevance in explaining correlations of financial returns \citep[e.g. see][]{barker2007geography,eckel2011measuring}. Notice also that it is unclear which geographical information could and should be used to analyse stock market data, especially if country-specific stocks are considered. For example, using the firms' headquarter to define spatial closeness can be reductive, considering that productive plants can be located in many alternative places. For this reason, many authors propose to use alternative definitions of similarity across stocks measured in an attribute space rather than the geographical one, e.g., the space spanned by financial indicators, balance sheet positions, or alternative indicators \citep[e.g. see][]{asgharian2013spatial,fernandez2012spatial,fulle2022spatial}. In this way, spillovers arise from stocks with similar characteristics, although not necessarily close in the geographical space. 

Therefore, using spatial econometric approaches in the financial domain requires a different definition of the proximity of locations and how spatial weights are determined. By substituting spatial contiguity with a broader concept of similarity, we can construct a financial network and model the volatility of a stock as a function of past values and the relationship with adjacent nodes in the network. Working with this kind of model is convenient because financial networks provide a suitable framework for understanding the propagation mechanisms of shocks occurring in the market. 

As shown by many authors \cite[e.g., see][]{diebold2014network,demiris2014epidemic,barigozzi2017network,vinciotti2019effect,betancourt2020modelling,liu2021dynamic,zhou2023dynamic}, the financial market is well represented by networks where stocks are the nodes, and the edges reflect the degree of similarity across them. We introduce the spillovers from adjacent nodes in an ARCH-like manner. The information from adjacent network nodes can be successfully used to model relationships across stock returns and volatility. The idea that better predictions can be achieved by incorporating network information attracted the interest of many researchers in the field \cite[e.g. see][]{wu2022price,huang2023grouped}, but we still know very little, especially about the form of interactions, and further studies are needed. Indeed, network modelling of returns and volatilities is a recent flourishing research area in finance. 

We contribute to this literature in three main directions. First of all, differently from previous papers \citep[e.g.][]{caporin2015proximity,zhou2020network,billio2021networks}, we propose the use of a network log-ARCH model for volatility forecasting. Specifically, we consider instantaneous network interactions in an ARCH-like structure. The volatility may immediately spill over to adjacent/similar stocks reflecting the simultaneity of investors' trading decisions. Furthermore, our proposed modelling approach shares the same relevant features of exponential volatility models. Secondly, we extend the dynamic spatiotemporal ARCH models of \cite{otto2022dynamic} with homogeneous temporal ARCH effects by introducing stock-specific temporal ARCH parameters. Thirdly, the forecasting performance is rigorously evaluated, which has not been the focus of prior studies on Network GARCH models \citep[e.g.][]{caporin2015proximity,billio2021networks,zhou2020network}. Therefore, we fill this gap by investigating the usefulness of introducing network structure in the model from a forecasting perspective. Fourthly, under the assumption of unknown locations, we evaluate how the approach used for constructing the network affects the forecasting results. More precisely, several network and edge weight definitions are compared. Previous papers proposed the construction of an adjacency based on subjective criteria such as the shared shareholders \cite{zhou2020network} or the industry sectors \cite{caporin2015proximity,billio2021networks}, which are not automatic data-driven procedures. Differently, we adopt three alternative definitions of similarity across stocks, taking inspiration from time series clustering literature \cite[for an overview, see][]{maharaj2019}.  In doing so, we adopt an intuitive and fast data-driven approach for constructing the network, which only requires the time series to be predicted for its construction. Then, we build the networks considering inverse-distance and $k$-nearest neighbours, obtaining twelve alternative Network log-ARCH models. Eventually, ensemble forecasts to combine all model alternatives are considered to further improve the out-of-sample forecasting performance.

The empirical experiment on stocks included in the Dow Jones Industrial Average Index demonstrates that the proposed network-based log-ARCH approach provides more accurate out-of-sample forecasts than the traditional log-GARCH modelling relying solely on temporal information. Moreover, considering predictive accuracy tests, we find that both the similarity measure and the procedure employed in constructing the network model affect the out-of-sample forecasting accuracy. Although all the network approaches outperform the benchmark, we identify a superior subset (three out of twelve) of network models with statistically more accurate forecasts. 

The rest of the paper is structured as follows. Section \ref{sec:theory} discusses the dynamic spatiotemporal log-ARCH model of \cite{otto2022dynamic} and how spatial weights are determined. Section 3 presents the data used for the empirical analysis and explains how the out-of-sample forecasting methodology is conducted. Section \ref{sec:results} shows the primary results: forecasting accuracy and predictive accuracy tests. Section \ref{sec:conclusion} concludes with final remarks and some suggestions for future research.

\section{Forecasting models}\label{sec:theory}

In the following sections, we will introduce the modelling framework used for forecasting comparisons. We consider that we observe a process of stock market returns on a network $G = (V, E)$, which consists of a set of nodes/vertices $V$ (i.e., stocks) that are possibly connected by directed or undirected edges. These edges are contained in the set $E$. Furthermore, we observe a process $\{Y_t(s_i) : t = 1, \ldots, T, s_i \in V \}$ of nodal attributes across time (i.e., return series). In particular, our focus will be on the volatility of this process, which should be forecasted for $T + 1, T + 2, \ldots$ using the dynamic network ARCH model. Moreover, let $V = \{s_1, \ldots, s_n\}$ be the set of $n$ stocks and $\boldsymbol{Y}_t = (Y_t(s_1), \ldots, Y_t(s_n))'$ be the $n$-dimensional vector of the observed process on the network. It is important to note that we do not consider dynamic networks (i.e., time-varying sets of nodes and/or edges), but the networks are assumed to be constant over time.

\subsection{Univariate logarithmic ARCH models (baseline model)}\label{sec:theory_logARCH}

We start our model comparison with univariate logarithmic ARCH (log-ARCH) models, which are fitted independently to each time series. In this way, we get a very flexible model allowing for different dependence structures for each stock to get the forecast performance of the volatility series. The model was originally proposed by \cite{geweke1986commet}. To be precise, the log-ARCH($P$) for an $i$-th stock can be written as follows
\begin{align}
    Y_{t}(s_i) &= \sqrt{h_{t}(s_i)} \varepsilon_{t}(s_i), \label{eq:logarch_Y}\\
    \ln h_{t}(s_i) &= \omega_i + \sum_{p=1}^{p} \gamma_{ip} \ln Y^2_{t-p}(s_i) \, , \label{eq:logarch_h}
\end{align}
where $\omega_i$ is the constant term, $\gamma_{i1}, \ldots, \gamma_{iP}$ are the ARCH parameters for the $i$-th stock, and $P$ is the order of the log-ARCH process. We can fit $n$ univariate log-ARCH models and predict stocks' volatilities by only using idiosyncratic temporal information. The $n(P+1)$ unknown parameters of the log-ARCH model can be consistently estimated via ARMA representation of the process \citep[e.g. see][]{sucarrat2016estimation}. Applying a log-square transformation of \eqref{eq:logarch_Y} shows that $\ln Y^2_{t}(s_i) = \ln h_{t}(s_i) + \ln \varepsilon^2_{t}(s_i)$. As a consequence,
\begin{equation*}
     \ln Y^2_{t}(s_i)
    = \omega_i + \sum_{p=1}^{P} \gamma_{ip} \ln Y^2_{t-p}(s_i) + \ln \varepsilon^2_{t}(s_i) \, .
\end{equation*}
However, since the distribution of $\ln \varepsilon^2_{t}(s_i)$ does not have a mean of zero, $E\left(\ln \varepsilon^2_{t}(s_i)\right)$ is added and subtracted from sides of the equation, leading to
\begin{equation}
\label{logarcharma}
  \ln Y^2_{t}(s_i) =\phi_{i0}+\sum_{p=1}^P \phi_{ip} \ln Y^2_{t-p}(s_i) + u_{t}(s_i)
\end{equation}
with a new error term $u_{t}(s_i)= \ln \varepsilon^2_{t}(s_i) - E\left(\ln \varepsilon^2_{t}(s_i)\right)$ and the constant $\phi_{i0} = \omega_i +  E\left(\ln \varepsilon^2_{t}(s_i)\right)$. Notice that $E\left(\ln \varepsilon^2_{t}(s_i)\right) = \mu_i^*$ for all $t$, so this transformation ensures that the error term $u_{t}(s_i)$ has a zero mean. Hence, the ARMA representation \eqref{logarcharma} allows for consistent estimation of the log-ARCH parameters. More precisely, we have that $\gamma_{ip} = \phi_{ip}$ and $\omega = \phi_{i0}  -  E\left(\ln \varepsilon^2_{t}(s_i)\right)$.

From the discussion highlighted so far, it is clear that we need an estimate of the term $E\left(\ln \varepsilon^2_{t}(s_i)\right)$ to estimate $\omega_i$ . \cite{bauwens2010general,sucarrat2012automated,sucarrat2016estimation} propose ex-post scale adjustments based on the estimated residuals $\{\hat{u}_{t}(s_i) , t = 1, \ldots, T\}$, i.e.,
\begin{equation}
\label{const}
   \hat{\mu}_i^* = - \ln \left[\frac{1}{T} \sum_{t=1}^{T} \exp(\hat{u}_{t}(s_i))\right] \, .
\end{equation}
It is a consistent and asymptotic normal smearing estimator of the log-square transformed errors' mean \citep[see][]{duan1983smearing,sucarrat2016estimation,francq2018exponential}. 

Lastly, the resulting the one-step-ahead forecast at $T+1$ of the log-ARCH(1) model is given by
\begin{equation}
  \ln \hat{h}_{t+1}(s_i) = \left[\hat{\phi}_{i0} + \ln \left(\frac{1}{T} \sum_{t=1}^{T} \exp(\hat{u}_{t}(s_i))\right)\right] + \hat{\phi}_{i1} \ln y^2_{t}(s_i) \, ,
\end{equation}
with $y^2_{t}(s_i)$ being the observed values of $Y^2_{t}(s_i)$. For the comparison, we will focus on one-step-ahead forecasts and a model order of $P = 1$, but the methods can easily be extended to multi-step-ahead forecasts and higher model orders. 

Generally, network processes could also be represented as multivariate time series. Thus, a natural extension to account for dependence between the nodes would be multivariate log-ARCH models \citep{francq2017equation}. However, they are limited in the sense that (1) they do not account for the inherent network structure, and (2) the number of parameters grows quadratically when the number of nodes $n$ increases. It is worth mentioning that the time series length (without any changes in the parameters or structural breaks) must be larger than $n^2$ to get unique and reasonably precise estimates. In this paper, we particularly focus on the case with large $n$ (compared to the length of the time series $T$). Hence, we propose to include instantaneous ARCH effects across the network to describe the dependence between the stock returns. In other words, the log volatility $\ln h_t(s_i)$ of the $i$-th stock is influenced by all other observations $y_t(s_j)$ for $j = 1, \ldots, n$ and $j \neq i$, whereby the dependence structure is determined by the edges $E$ of the network. In this way, large volatilities (i.e., large values of $y_t(s_j)^2$) can spill over to the adjacent nodes and lead to an increase in $\ln h_t(s_i)$. As a consequence, volatility clusters can be observed across the network.

\subsection{Dynamic network logarithmic ARCH model}\label{sec:theory_networkARCH}

The new dynamic network log-ARCH model is based on dynamic spatiotemporal ARCH models proposed by \cite{otto2022dynamic}. As for the univariate log-ARCH models, the observed process is given by
\begin{align}
    Y_{t}(s_i) & = \sqrt{h_{t}(s_i)} \varepsilon_{t}(s_i), \label{eq:network_arch_Y}
\end{align}
but now $h_{t}(s_i)$ is being influenced by past observations at the same node, $Y_{t-1}(s_i)$, and simultaneously by the adjacent observations at the same time point, $\{Y_{t}(s_j) : j \in E_i \}$, where $E_i$ is the subset of edges with links to node $s_i$. Let $\boldsymbol{h}^*_t = (\ln h^2_t(s_1), \ldots, \ln h^2_t(s_n))'$ and $\boldsymbol{Y}^*_t = (\ln Y^2_t(s_1), \ldots, \ln Y^2_t(s_n))'$. Then, the network log-ARCH process of order one can be written as follows
\begin{align}
\boldsymbol{h}^*_t =  \boldsymbol{\omega} + \mathbf{\Gamma} \boldsymbol{Y}^*_{t-1} + \rho \mathbf{W} \boldsymbol{Y}^*_t  \, , \label{eq:network_arch_h}
\end{align}
where $\mathbf{W} = (w_{ij})_{i,j = 1, \ldots, n}$ is a matrix of edge weights, which define the relative degree of the volatility spillovers, $\rho$ is an unknown parameter for these instantaneous network interactions, $\mathbf{\Gamma} = \text{diag}(\gamma_1, \ldots, \gamma_n)'$ is a diagonal matrix of stock-specific temporal ARCH effects, and $\boldsymbol{\omega} = (\omega_1, \ldots, \omega_n)'$ is the constant term. The matrix of edge weights $\mathbf{W}$ is analogously specified as the spatial weight matrix in spatial econometrics. That is, the diagonal entries are supposed to be zero (i.e., no self-loops), the matrix is non-stochastic, and uniformly bounded in row and column sums in absolute terms. The latter assumption is needed to limit the network interactions to a constant degree when the number of nodes $n$ is increasing. Typical choices are inverse-distance matrices (e.g., for road networks) or $k$-nearest neighbours matrices, where the proximity is defined by any network characteristic. We will discuss the definition of $\mathbf{W}$ in more detail in Section \ref{sec:weights}.

The log-volatility terms follow a process characterized by the presence of instantaneous network effects, which is the key difference to previously proposed network GARCH models \citep[e.g.][]{zhou2020network} or GARCH models including artificial neural network structures \citep[e.g.][]{donaldson1997artificial,kristjanpoller2015gold}. These previous models include network interactions only at the first temporal lag. That is, network spillovers can only happen in the next time instance but not instantaneously. The network log-ARCH model allows for deriving an ARMA representation of the model, namely
\begin{align}\label{eq:networkARMA}
\boldsymbol{Y}^{*}_t = \boldsymbol{\phi}_0 + \rho \mathbf{W}\boldsymbol{Y}^{*}_t +  \mathbf{\Gamma} \boldsymbol{Y}^{*}_{t-1} + \boldsymbol{u}_t,
\end{align}
where $\boldsymbol{u}_t = (\ln \varepsilon^2_t(s_1) - E(\ln \varepsilon^2_t(s_1)), \ldots, \ln \varepsilon^2_t(s_n) - E(\ln \varepsilon^2_t(s_n)))'$ are the log-squared errors and $\boldsymbol{\phi}_0 = \boldsymbol{\omega} + \boldsymbol{\mu}^*$ with $\boldsymbol{\mu}^* = \left(E(\ln \varepsilon^2_t(s_1)), \ldots, E(\ln \varepsilon^2_t(s_n))\right)'$. It is important to note that we allow for different means $E(\ln \varepsilon^2_t(s_i))$ for each stock, which is constant over time. To estimate $\boldsymbol{\phi}_0^*$, we propose to use the smearing estimate as proposed by \cite{sucarrat2016estimation} for time-series log-ARCH models, i.e.,
\begin{align}\label{eq:constant_estimation}
\hat{\boldsymbol{\phi}}_0^* = \frac{1}{T}\sum_{t = 1}^{T} \hat{\boldsymbol{u}}_t .
\end{align}
From \eqref{eq:networkARMA}, we see that the instantaneous spillovers lead to endogeneity in the model (i.e., $\boldsymbol{Y}^{*}_t$ appears on both sides of the equation), which is not the case when the network interactions are restricted to the first temporal lag like in \cite{zhou2020network}. Hence, we apply the estimation proposed by \cite{otto2022dynamic} based on orthonormal transformations and the generalised method of moments (GMM). The key idea of the GMM estimator is to instrument $\mathbf{W}\boldsymbol{Y}^{*}_t$ by higher-order network lags originally proposed for spatiotemporal autoregressive models by \cite{Lee:2007,Lee:2014}. For further details, we refer the interested reader to \cite{otto2022dynamic}.

Finally, we obtain the one-step-ahead forecast of all stocks at time $T+1$ as
\begin{equation}
\boldsymbol{h}^*_{T+1} = \left( \mathbf{I}_n - \hat{\rho} \mathbf{W} \right)^{-1} \left[ \hat{\mathbf{\Gamma}} \boldsymbol{Y}^{*}_T + \hat{\boldsymbol{\phi}}_0 \right] \, ,
\end{equation}
where $\mathbf{I}_n$ is the $n$-dimensional identity matrix. Notice that $\boldsymbol{\phi}_0$ includes $\boldsymbol{\omega}$ and $\boldsymbol{\mu}^*$, but these two quantities are jointly estimated from the residuals' process as in \eqref{eq:constant_estimation} because the orthonormal transformation eliminates all cross-sectional fixed effects.

For this network log-ARCH model, finding a suitable edge weight matrix for the ARCH-type interactions across the network is crucial. The edges are typically unknown for financial networks or stock market interactions and therefore have to be estimated. Nevertheless, it is reasonable to assume that with the increasing similarity between the stocks, they are more likely to experience spillovers in the risks, i.e., conditional volatilities. In the following section, we discuss several options to estimate the similarity/dissimilarity in the stock return series, which is a basis for the edge weights in $\mathbf{W}$.

\subsection{Determining similarity across stocks}\label{sec:weights}

Measuring distance across time series can be done in many different ways. In this regard, one mainly distinguishes between raw data-based, feature-based and model-based approaches \citep{maharaj2019}. In the first class, we consider dissimilarities computed on raw data, such as using standard Euclidean distance on temporal observations or Dynamic Time Warping when stocks have different lengths. Following a future-based approach, dissimilarities across financial time series can be calculated based on asset correlations \citep{mantegna1999hierarchical,tumminello2010correlation}, auto-correlation structures \citep{d2009autocorrelation}, periodograms \citep{caiado2006periodogram,caiado2020fragmented} or Hurst exponents \citep{lahmiri2016clustering,cerqueti2023fuzzy}. Further, model-based approaches measure dissimilarity across stocks by considering parameters estimated from statistical models, such as ARIMA \citep{piccolo1990distance}, GARCH \citep[e.g. see][]{otranto2008clustering,d2016garch}, Multiplicative Error Model \citep[e.g. see][]{gallo2021classifying}, or score-driven models \citep[e.g. see][]{cerqueti2021model,cerqueti2022weighted}.  The most commonly employed approaches are those based on volatility dynamics when dealing with stock returns. Indeed, volatility-based approaches allow measuring similarities by directly exploiting conditional heteroscedasticity. 

In this paper, we consider three alternative configurations for the matrix $\mathbf{W}$, that is, the standard Euclidean distance across stock returns, the use of a correlation-based approach as suggested in \cite{mantegna1999hierarchical} and a model-based approach based on the log-ARCH estimates. For the first approach, the dissimilarities over time are computed as follows
\begin{equation}
\label{dist1}
    d_{ij} = \sqrt{\sum_{t=1}^{T} \left( y_{t}(s_i) - y_{t}(s_j) \right)^2} \, .
\end{equation}
Considering the correlation-based approach, instead, the generic entries of the dissimilarity matrix are given by
\begin{equation}
\label{dist2}
    d_{ij} = \sqrt{2 \left( 1- \rho_{ij} \right)}
\end{equation}
with $\rho_{ij}$ being the estimated correlation coefficient between the stocks $i$ and $j$ over the entire time horizon $t = 1, \ldots, T$. In this way, we assume that stocks are similar to each other if their correlation is high. Finally, we propose using a log-ARCH approach, accounting for the stocks' underlying log-volatility dynamics. According to \cite{piccolo1990distance}, we can define the dissimilarity between two time series $i$ and $j$ by means of the AR$(\infty)$ of log-squared returns, i.e.,
\begin{equation}
\label{dist3}
    d_{ij} = \sqrt{\sum_{p=1}^{\infty} \left( \gamma_{ip} - \gamma_{jp} \right)^2},
\end{equation}
where $\gamma_{ip}$ is the autoregressive parameter (of order $p$)log-squared returns series of the $i$-th stock. In practice, the infinite sum in \eqref{dist3} has to be truncated at some order $P$. The selection can be made according to the AIC or BIC information criteria. If the two time series $i$ and $j$ have different orders, $P_i$ and $P_j$, we take $P=\max(P_1, P_2)$ and let $\gamma_{ip} = 0$ for $P > P_1 $ and, similarly, $\gamma_{jp} = 0$ for $P > P_2$ \citep{piccolo1990distance}. 

Based on the pairwise distances in \eqref{dist1} to \eqref{dist3}, we consider two different strategies for constructing the weighting matrix $\mathbf{W}$. First, $\mathbf{W}$ is defined as inverse-distance weight matrix with
\begin{equation}
w_{ij} = d_{ij}^{-1} \qquad \text{for all $i,j = 1,\ldots,n$} \, . \label{eq:invdist}
\end{equation} 
Secondly, $\mathbf{W}$ is considered to have constant edge weights based on the $k$-nearest neighbours, i.e., 
\begin{equation}
w_{ij} = \left\{
\begin{array}{cc}
    1/k & \text{if $s_j$ is closer than the $k+1$} \\
    & \text{nearest neighbours of $s_i$} \\
    0   & \text{otherwise.}
\end{array}
\right. \label{eq:knn}
\end{equation} 
The so-obtained weighting matrix $\mathbf{W}$ expresses distances in the (temporal) attribute space. Whereas the inverse-distance matrix is a symmetric and dense matrix by construction with equal weights for both directions (i.e., undirected graph, fully connected), the $k$-nearest-neighbours matrix is not necessarily symmetric (it will be usually asymmetric), but each neighbour has the same weight (i.e., directed graph, homogeneously weighted). Alternatively, the weights could be obtained from estimated distances in balance sheet data \citep[cf.][]{fulle2022spatial}. 

\section{Empirical analysis: data and forecasting methodology}\label{sec:results}

For our empirical experiment, we focused on volatility forecasting of stocks in the Dow Jones Industrial Average Index. To facilitate replication and further analysis, we have made our code, data, and examples available at \url{philot789.github.io/Network_ARCH/}.

\subsection{Data}

\begin{figure}
         \centering
         \includegraphics[width=0.8\linewidth]{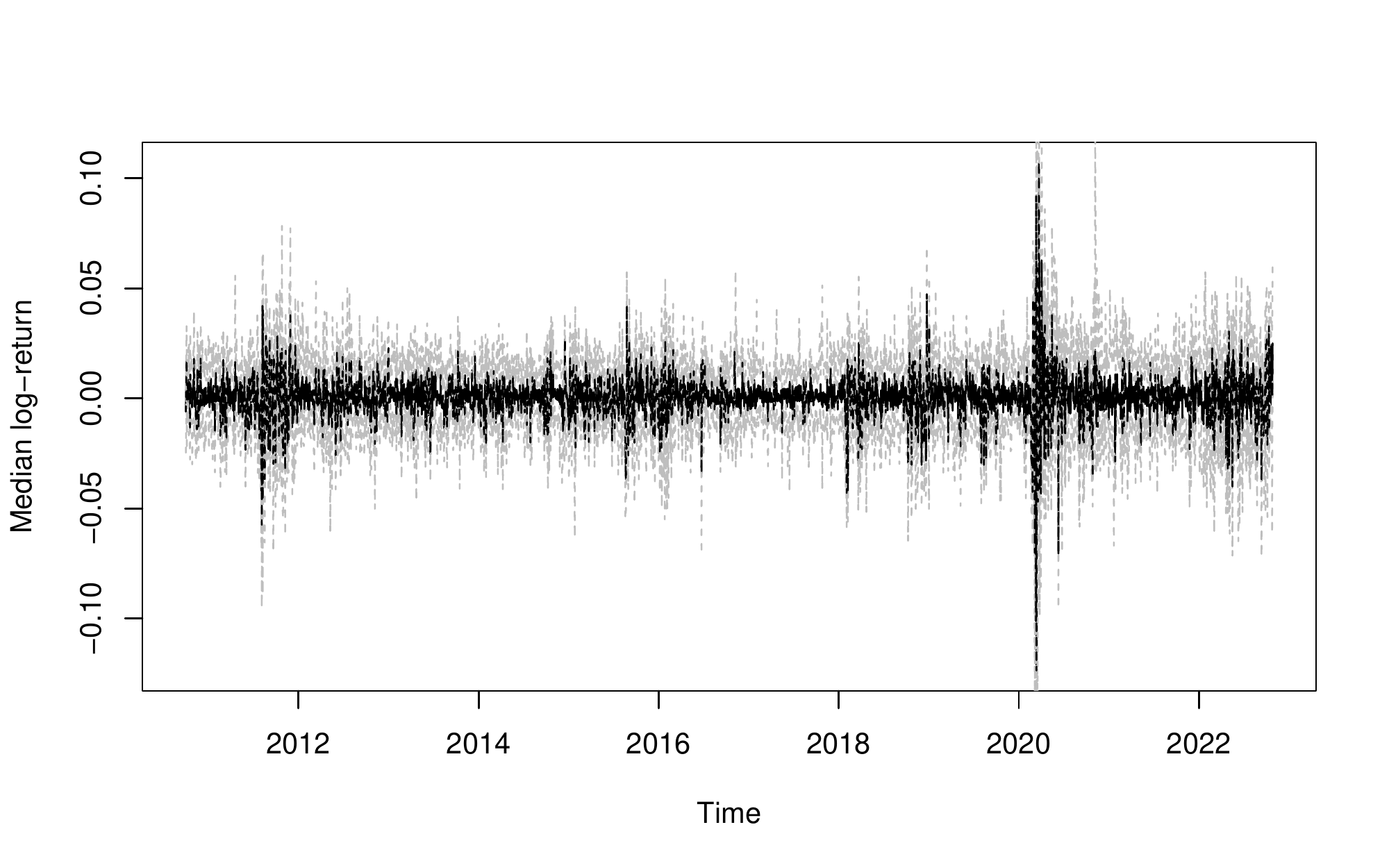}\\
         \includegraphics[width=0.8\linewidth]{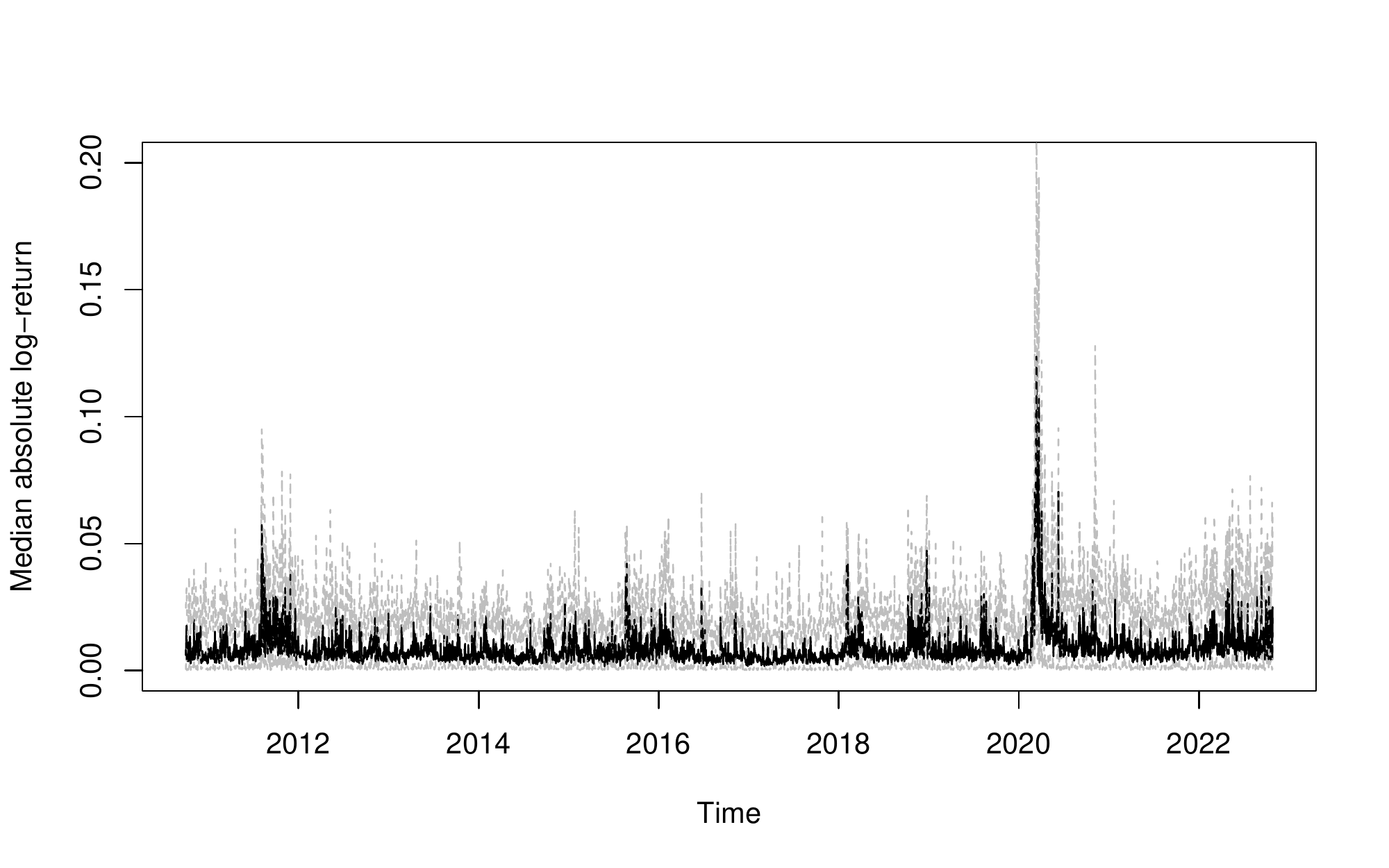}
        \caption{Summary of $n = 29$ time series of log-returns. Top: Median log-returns of each day (solid line) and the 5\% and 95\% quantiles (dashed lines), bottom: Median (solid line) and the 5\% and 95\% quantiles of the absolute log-returns to depict the temporally varying volatility.}  
         \label{fig:timeseries}
\end{figure}

The time series of daily stock returns span from October 1, 2010, to October 31, 2022. We removed any stocks with missing values from our analysis. The list of considered stocks is shown in Tab. \ref{tab:stocks} along with main descriptive statistics. The return series are displayed in Fig. \ref{fig:timeseries} in terms of the median log-returns of all stocks on each day (first row), and their median absolute returns as a measure of the stock market volatility (second row). In total, we observe the logarithmic returns of $n = 29$ nodes/stock for $T = 3040$ days. To evaluate the benefit of including instantaneous network ARCH effects, we include the same information for all predictions, i.e., we choose the temporal lag order $P = 1$ for both models. We allow for different values of $\mu_i^*$ for each stock and different temporal ARCH parameters for each stock. That is, for $\rho = 1$, the network model is identical to the independent time-series log-ARCH models considered as benchmark model. It is worth noting that the number of stocks $n$ is large compared to the length of the time series, such that the number of parameters of multivariate GARCH would exceed a reasonable degree and it is useful to include a certain structure of the covariance. Here, we consider the proposed network ARCH structure. For this reason, we compare the forecasting performance for various choices of edge weights, namely all three distance measures described in \eqref{dist1}-\eqref{dist3} and inverse-distance weighting (models A.1, A.2, A.3) and $k$-nearest neighbours weights with $k = 2, 3, 5$, and $10$ (models B.$k$.1, B.$k$.2, and B.$k$.3).

\begin{figure}
    \centering
    \includegraphics[width=0.42\linewidth]{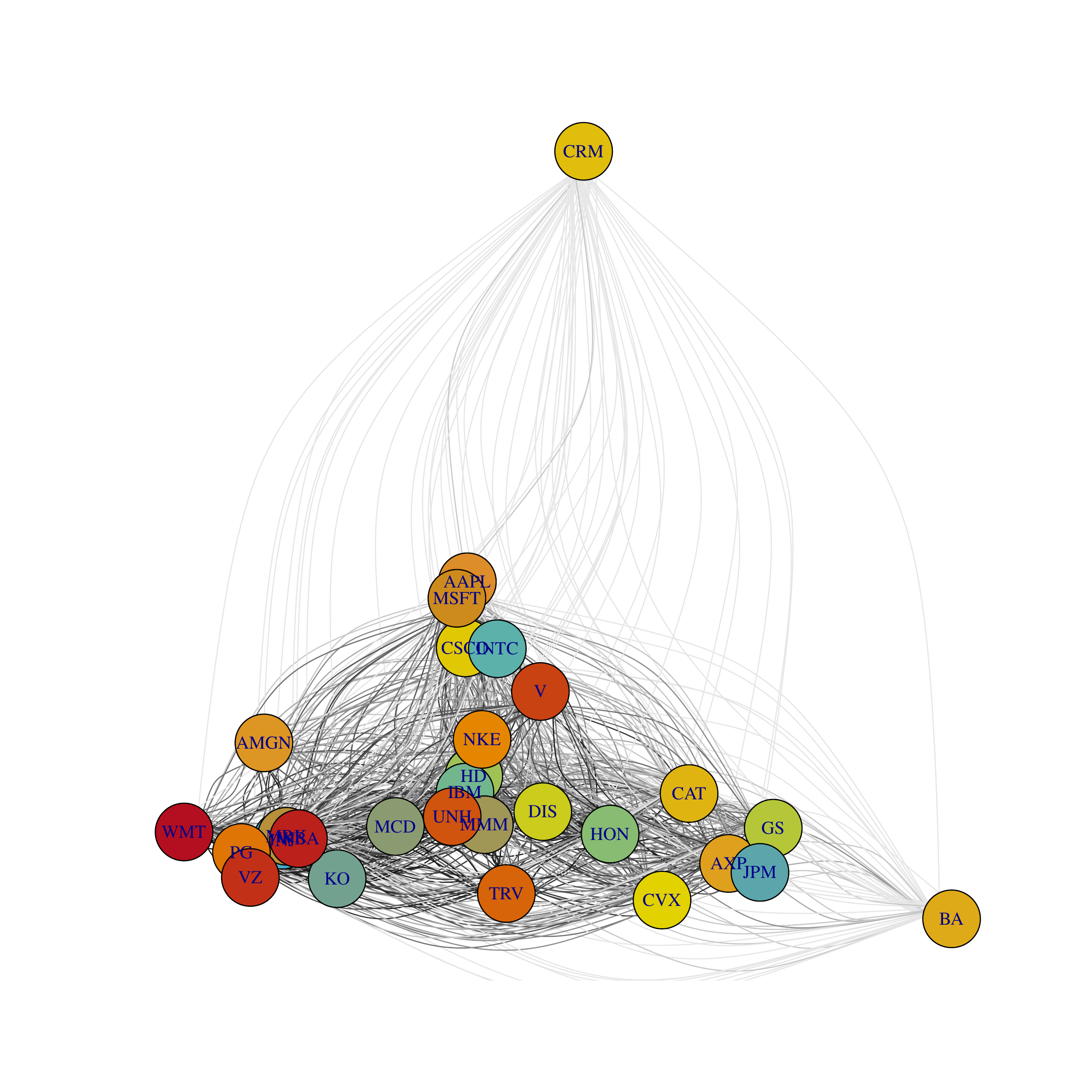}
    \includegraphics[width=0.42\linewidth]{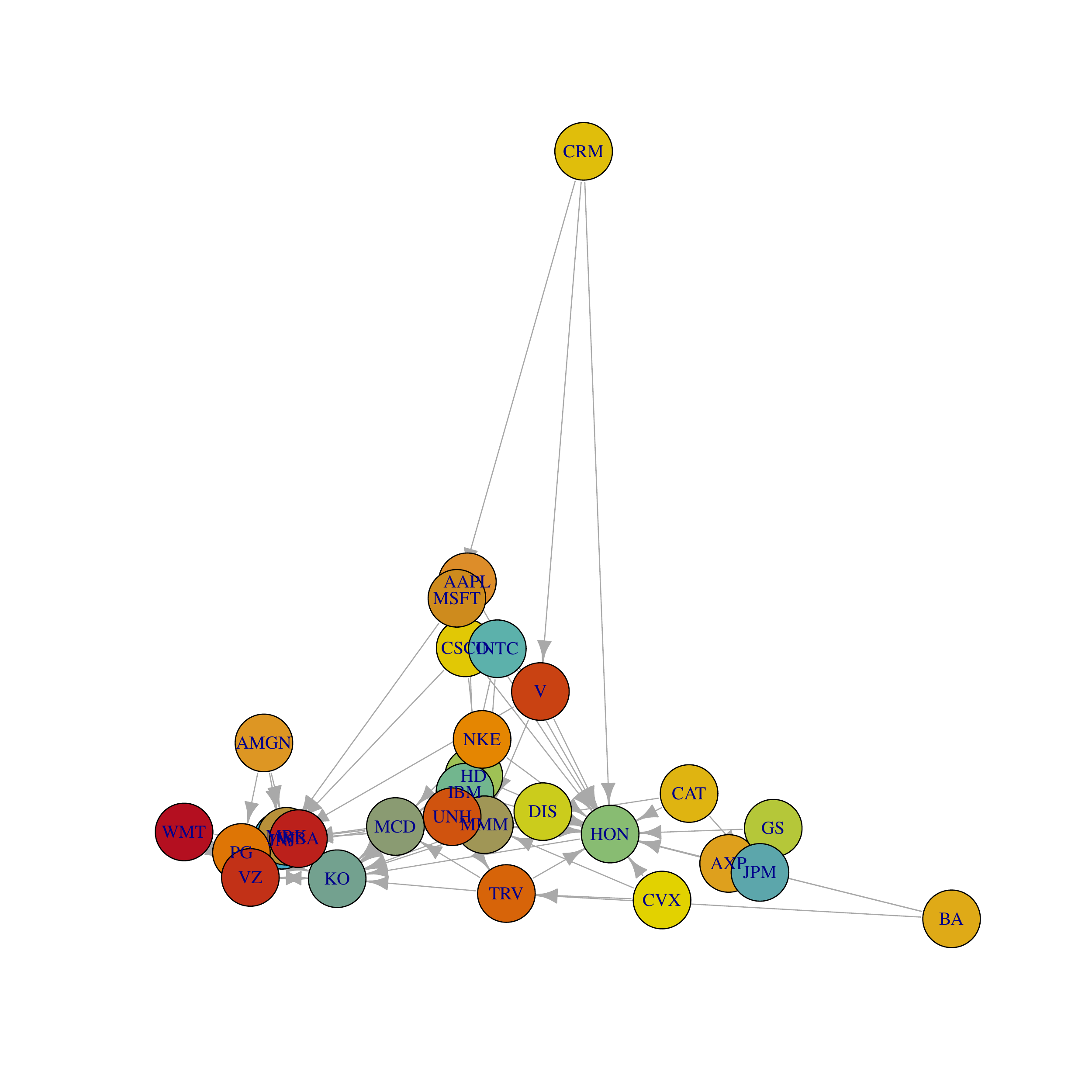}\\
    \includegraphics[width=0.42\linewidth]{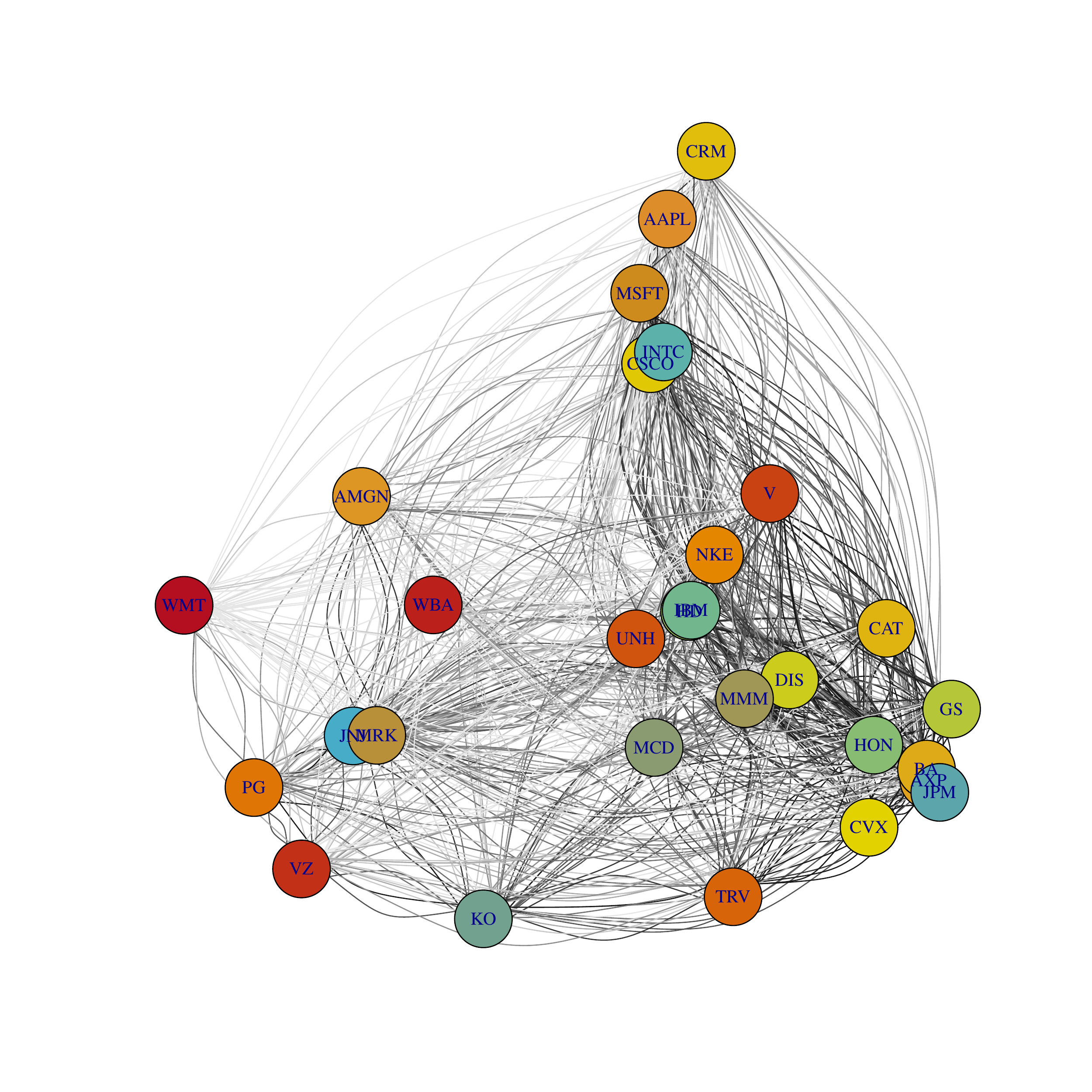}
    \includegraphics[width=0.42\linewidth]{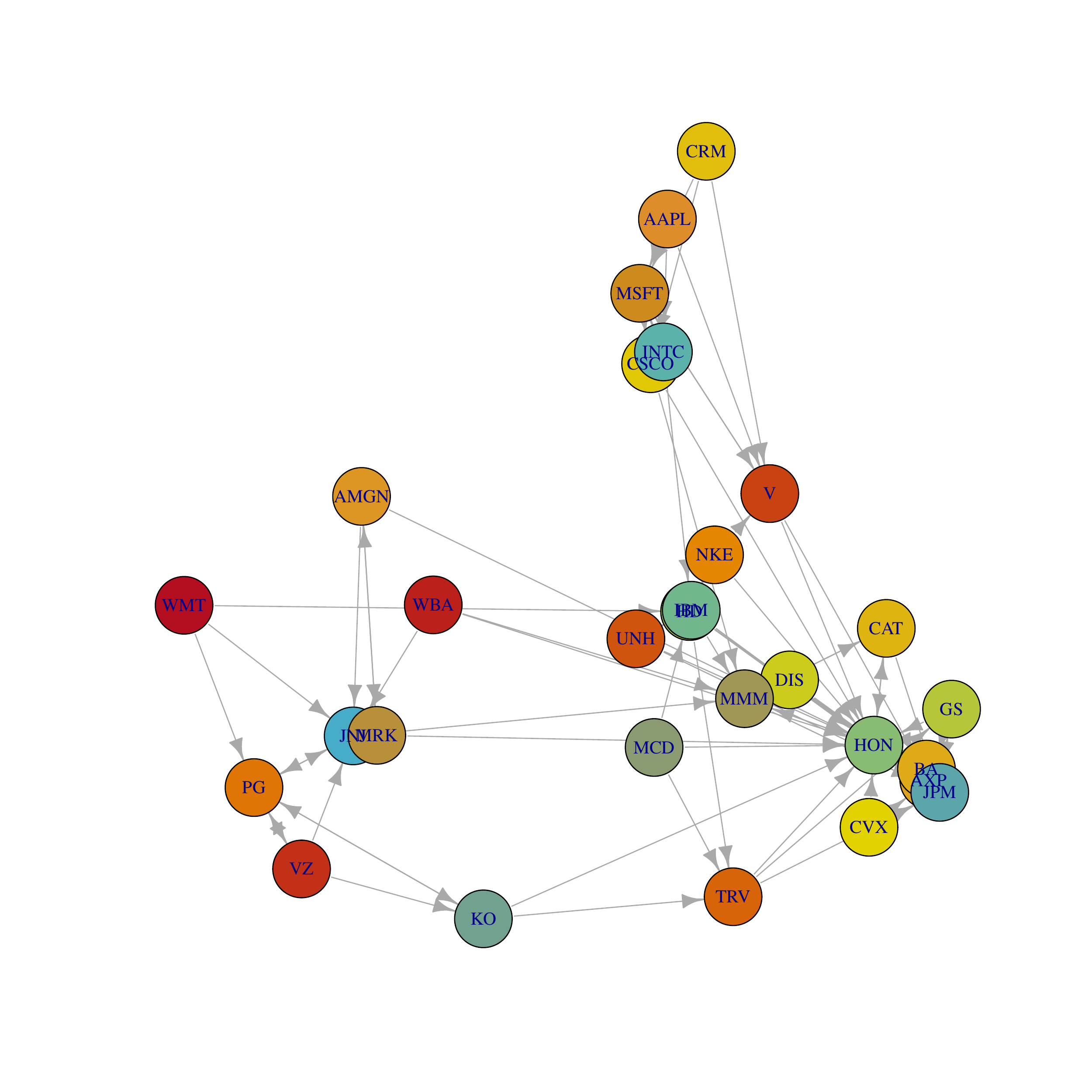}\\
    \includegraphics[width=0.42\linewidth]{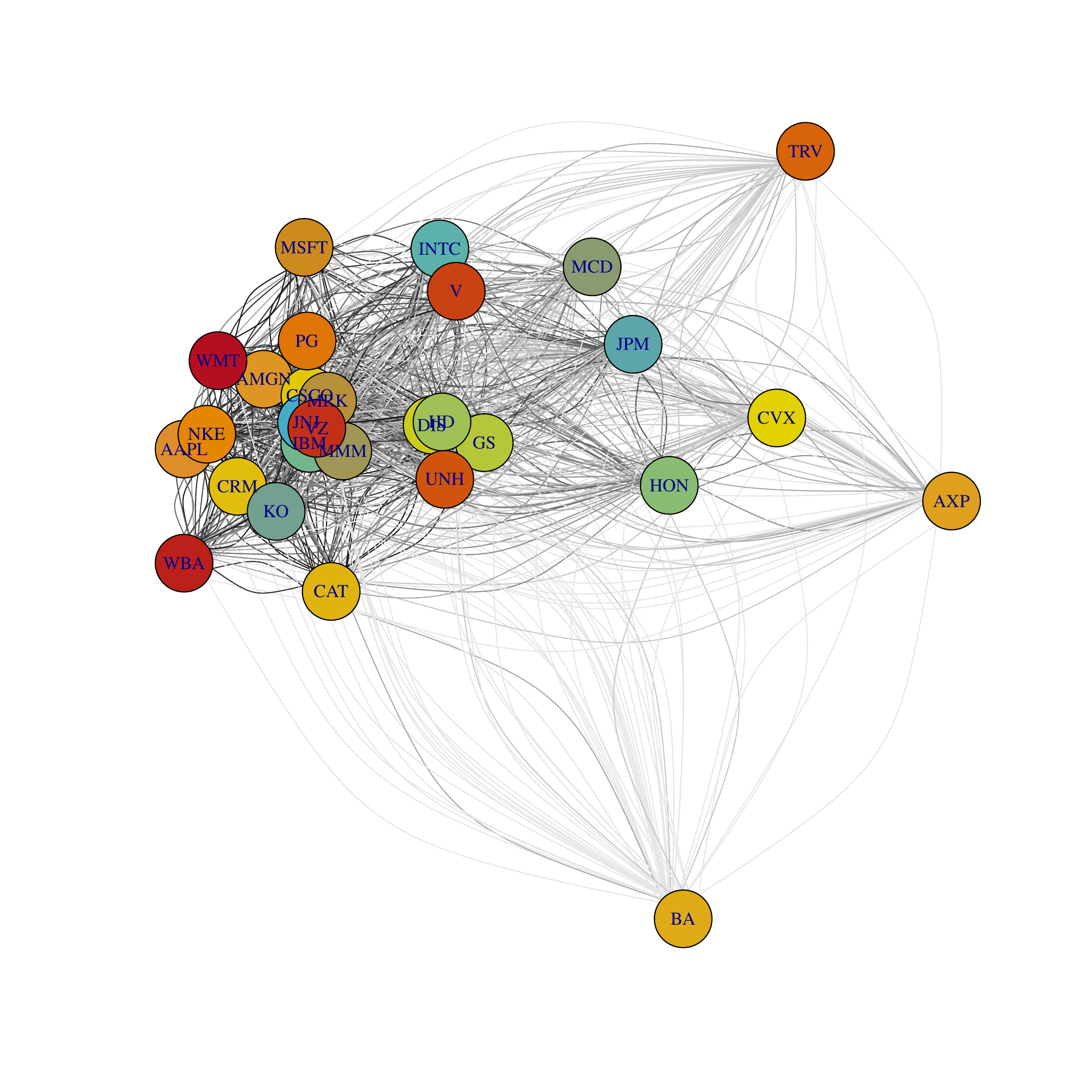}
    \includegraphics[width=0.42\linewidth]{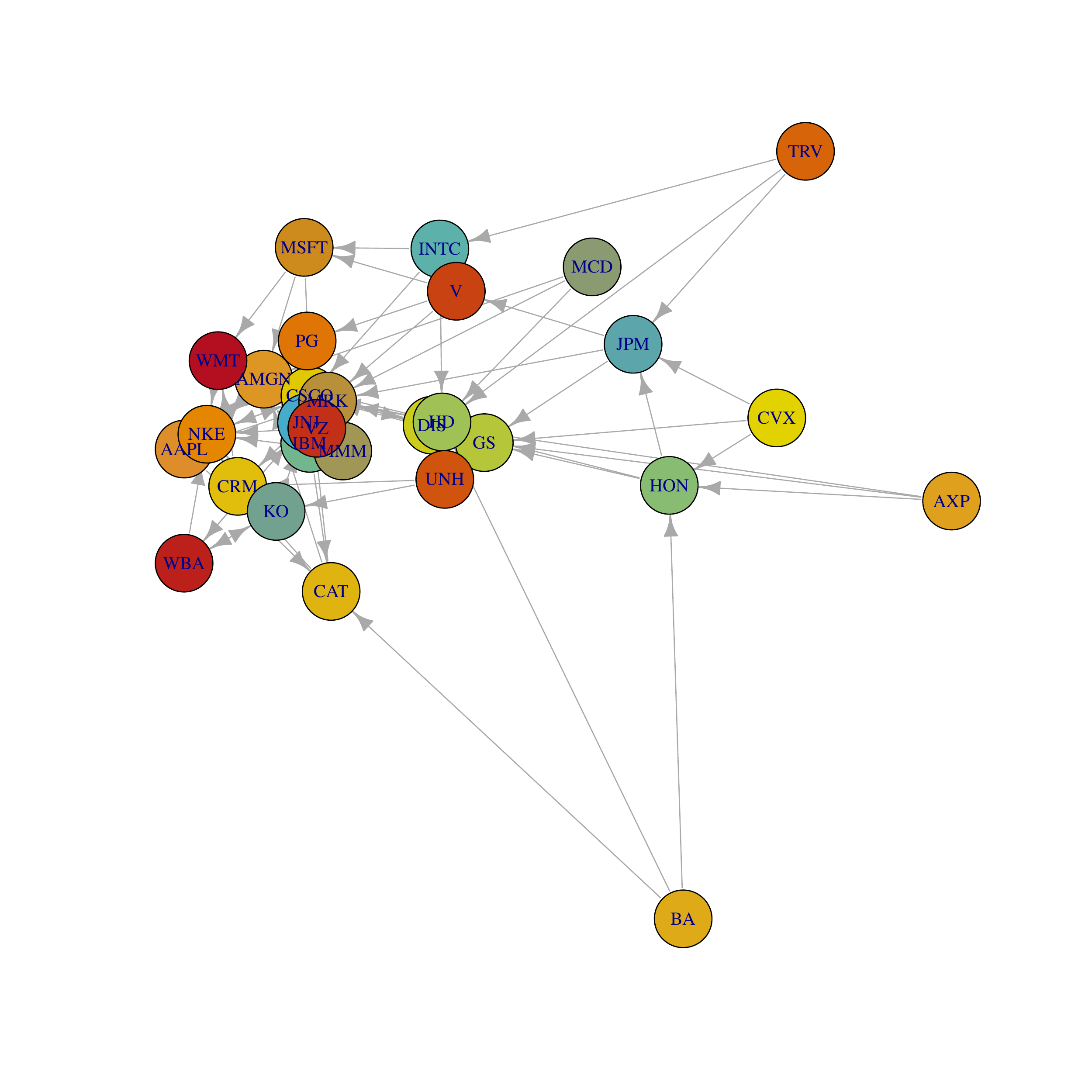}
    \caption{Network of the considered stock with different dissimilarity measures and weighting schemes. The nodes are located according to their distances. Top row: Euclidean dissimilarity \eqref{dist1}, center row: correlation dissimilarity \eqref{dist2}, bottom row: log-ARCH dissimilarity \eqref{dist3}. Left column: inverse-distance edge weights \eqref{eq:invdist} with edges coloured according to their weights (i.e., the higher the weight, the darker the edge), right column: $k = 3$-nearest-neighbours weights \eqref{eq:knn}, where the arrows point towards the direction of the influence.}
    \label{fig:net1}
\end{figure}

The networks obtained considering the alternative approaches are shown in Fig. \ref{fig:net1}. The nodes in Fig. \ref{fig:net1} are located according to their distances, and, in the case of the inverse distance approach, the edges are coloured according to their weights such that the higher the weight, the darker the edge. The network structures highlight interesting differences which can be exploited in the forecasting task.

In the case of Euclidean distances as described in \eqref{dist1}, we observe two outlying nodes, namely the CRM and BA stocks. When the weights are constructed based on the inverse distance (left-hand graphs), these two stocks have a minor influence on all other stocks. By contrast, for the $k$-nearest neighbours weights (right-hand graphs), their influence to adjacent stocks will be of the same degree. Interestingly, considering the main descriptive statistics in Tab. \ref{tab:stocks}, BA and CRM have the highest variability in the sample. Specifically, BA has the lowest minimum value, while CRM has the highest maximum. Thus, we observe the highest Euclidean distance for these two stocks, which is not robust in the case of outlying observations. Furthermore, these two stocks show the most volatile patterns in terms of their returns. However, under this network structure and inverse-distance weights, the information included of these two stocks will have a pretty low relevance in forecasting other stocks in the network. Vice-versa, the leading network structure is characterised by a central block (comprising, among the others, IBM, UNH, and MMM) affected mainly by the influence of the other stocks. Simultaneously, this group of stocks strongly affect those placed in the tails of the network. The central block of stocks has a stronger relationship with CRM, while the effect of BA is more pronounced for stocks in the right tail of the network.

The second row of Fig. \ref{fig:net1} shows the networks constructed under correlation distance defined in \eqref{dist2}. From one side, under this scenario, the stock BA is not far away from the other stocks, as it is highly correlated with AXP, HON, JPM and the other closer stocks. The same pattern was also observed for the Euclidean distance. On the other side, CRM is still the most distant from the others, although it shows an interestingly high correlation with IT-related stocks, such as AAPL, MSFT, INTC, and CSCO. In light of this evidence, stocks belonging to the IT-related sectors would rely on information from stocks of the same or similar industry sector. In this way, the information included in highly correlated stocks is successfully employed to improve the volatility forecasts due to the network structure. For the inverse-distance weights (left-hand graphs), this information spillover is represented by the edge colour, as the more relevant arrows have darker colours, and we observe a block of closely connected stocks in the bottom right area of the graph. Contrary to these nodes, stocks placed on the left show lighter arrows, meaning that their information would scarcely be used for predicting stocks in the right of the structure as well as their predictions rely mainly on idiosyncratic temporal information rather than the information coming from the network.

The last row of Fig. \ref{fig:net1} shows the network structure constructed in terms of the volatility-based measure given by \eqref{eq:invdist}. Under this network structure, closer stocks are those sharing similar log-volatility dynamics. BA is again an outlying stock in this case, as it was under the Euclidean-based network. Thus, information from BA will be scarcely used for predicting the other stocks, and, at the same time, BA forecasts will be mainly based on their temporally lagged values. The other two stocks are placed far from the main cluster are AXP and TRV. However, the edges of these stocks show darker colours compared with BA, meaning that these stocks still have informative power for the volatility dynamics of the network.

\begin{table}
\centering
\caption{Considered stocks in the empirical analysis and main descriptive statistics.}
\label{tab:stocks}
\scalebox{0.8}{\begin{tabular}{llrrrr}
  \hline
Company & Symbol & Mean & St. Dev. & Min & Max \\ 
  \hline
Apple & AAPL & 0.0010 & 0.0180 & -0.1377 & 0.1132 \\ 
  Amgen & AMGN & 0.0006 & 0.0153 & -0.1008 & 0.1034 \\ 
  American Express & AXP & 0.0005 & 0.0183 & -0.1604 & 0.1979 \\ 
  Boeing & BA & 0.0003 & 0.0232 & -0.2724 & 0.2177 \\ 
  Caterpillar & CAT & 0.0004 & 0.0183 & -0.1541 & 0.0983 \\ 
  Salesforce & CRM & 0.0006 & 0.0226 & -0.1730 & 0.2315 \\ 
  Cisco & CSCO & 0.0004 & 0.0169 & -0.1769 & 0.1480 \\ 
  Chevron & CVX & 0.0004 & 0.0177 & -0.2501 & 0.2049 \\ 
  Dow & DOW & 0.0004 & 0.0161 & -0.1391 & 0.1346 \\ 
  Goldman Sachs & GS & 0.0003 & 0.0182 & -0.1359 & 0.1620 \\ 
  Home Depot & HD & 0.0008 & 0.0148 & -0.2206 & 0.1288 \\ 
  Honeywell & HON & 0.0006 & 0.0147 & -0.1288 & 0.1404 \\ 
  IBM & IBM & 0.0002 & 0.0144 & -0.1375 & 0.1071 \\ 
  Intel & INTC & 0.0003 & 0.0187 & -0.1990 & 0.1783 \\ 
  Johnson \& Johnson & JNJ & 0.0005 & 0.0107 & -0.1058 & 0.0769 \\ 
  JPMorgan Chase & JPM & 0.0005 & 0.0179 & -0.1621 & 0.1656 \\ 
  Coca-Cola & KO & 0.0004 & 0.0111 & -0.1017 & 0.0628 \\ 
  McDonald's & MCD & 0.0005 & 0.0121 & -0.1729 & 0.1666 \\ 
  3M & MMM & 0.0002 & 0.0139 & -0.1386 & 0.1187 \\ 
  Merck & MRK & 0.0005 & 0.0131 & -0.1038 & 0.0990 \\ 
  Microsoft & MSFT & 0.0008 & 0.0164 & -0.1595 & 0.1329 \\ 
  Nike & NKE & 0.0006 & 0.0172 & -0.1371 & 0.1444 \\ 
  Procter \& Gamble & PG & 0.0004 & 0.0111 & -0.0914 & 0.1134 \\ 
  Travelers & TRV & 0.0005 & 0.0144 & -0.2332 & 0.1248 \\ 
  UnitedHealth & UNH & 0.0010 & 0.0161 & -0.1897 & 0.1204 \\ 
  Visa & V & 0.0008 & 0.0161 & -0.1456 & 0.1397 \\ 
  Verizon & VZ & 0.0002 & 0.0112 & -0.0697 & 0.0740 \\ 
  Walgreens Boots Alliance & WBA & 0.0001 & 0.0178 & -0.1548 & 0.1187 \\ 
  Walmart & WMT & 0.0004 & 0.0124 & -0.1208 & 0.1107 \\ 
   \hline
\end{tabular}}
\end{table}

\subsection{Forecasting evaluation}

The forecasting methodology is based on a rolling window procedure. For this reason, we first divide the sample into training and testing sets for each stock, leaving the last 500 observations (i.e., the last two years) for out-of-sample testing. The first $M=2540$ observations are used to obtain the edge weight matrix $\mathbf{W}$ and to estimate the model parameters. Then, the models are used for the one-step-ahead forecasts at $M+1$. Then, according to a rolling window procedure, the oldest observation is removed for the next step, and the new realised observation at $M+1$ is included in the estimation sample. Parameters are re-estimated with the new data, and the forecasts are obtained for $M+2$. This procedure is repeated until no new observation is available and all $T-M = 500$ volatilities were predicted for each stock. Thus, we always have an estimation window equal to 2540 observations at each recursion step.

To evaluate forecasting accuracy, we rely on two commonly employed accuracy metrics, namely the Root Mean Squared Forecast Error (RMSFE), i.e.,
\begin{equation}
    \text{RMSFE} = \sqrt{\frac{1}{T-M} \sum_{t=M+1}^{T} \left( \ln \hat{h}_{it} - \ln y^2_{it}\right)^2},
\end{equation}
and the Mean Absolute Forecast Error (MAFE), i.e.,
\begin{equation}
   \text{MAFE} =  \frac{1}{T-M} \sum_{t=M+1}^{T} \left| \ln \hat{h}_{it} - \ln y^2_{it}\right| .
\end{equation}
Notice that we use realised squared log returns as the proxy of volatility for the out-of-sample accuracy evaluation. Furthermore, we evaluate if the forecasting errors of the competing statistical models are significantly different by means of a predictive accuracy test.

Let $d_t = g \left( e_{1,t} \right) - g \left(e_{2,t}\right)$ be the error differential between two forecasting approaches up to some transformation $g(\cdot)$, that in this paper are squaring $g(e_{1,t}) = e^2_{1,t}$ and absolute value $g(e_{1,t}) = |e_{1,t}|$. Note that we performed the test independently for each stock; thus, we drop the index $i$ in this notation. Assuming covariance stationarity of the loss differential series $d_t$, \cite{diebold2002comparing} show that the sample mean of the loss differential
\begin{equation}
    \bar{d} \equiv \frac{1}{T-M} \sum_{t=M+1}^{T} d_t
\end{equation}
asymptotically follows a standard normal distribution. Hence, as test decision about the null hypothesis of equal forecast accuracies can be obtained based on the following statistic
\begin{equation}
\label{dmstat}
    \mathrm{DM}=\frac{\bar{d}}{\sqrt{\hat{V}(\bar{d})}}, \quad \mathcal{N} \sim \left(0,\hat{V}(\bar{d})\right) ,
\end{equation}
where $V(\bar{d})$ can be consistently estimated assuming a particular autocorrelation structure of the forecasting errors. To compare forecasts obtained from multiple models, we consider the Model Confidence Set (MCS) procedure of \cite{hansen2011model} for sequential testing based on the statistics in \eqref{dmstat}.

\section{Results of the out-of-sample forecasting experiment}

Below, we discuss the results of the out-of-sample experiment. We aim at predicting log volatility for the $n = 29$ stocks considered in Tab. \ref{tab:stocks}. Forecasts obtained from 13 alternative models are compared. The time-series log-ARCH is the selected benchmark, as it exploits temporal information only. For the network log-ARCH models, we considered three different distance measures and four alternative weight definitions (i.e., inverse-distance, $k=3$-nn, $k=5$-nn, $k=10$-nn), resulting on 12 different network models.

\subsection{Does network-based approach improves the forecasting accuracy?}
 
In what follows, we evaluate if the additional information from the network nodes is useful in forecasting volatility. An overview of the results in terms of both MAFE and RMSFE are shown in Tab. \ref{tab:weightcomparison}. The first column shows the log-ARCH results, while the other columns report the forecasting results of different network structures. Average RMSFE and MAFE values across stocks are reported for readers' convenience. Furthermore, we also report the ``best case'' of the benchmark, i.e., the stock which could be predicted the best in the benchmark model, which is the TRV stock. Likewise, the stock with the best predictions of the network models is reported, which is also the TRV regarding RMSFE and IBM regarding MAFE. The ``worst cases'' are MRK and VZ for the benchmark and the network models, respectively. In this way, we assess the performance of the network compared to the benchmark model in terms of averages and considering both best and worse scenarios. Interestingly, all the network-based models provide more accurate out-of-sample predictions compared to the log-ARCH, because of their increased model flexibility. 

Let us consider the RMSFE results first. The log-ARCH model provides an average RSME of 2.82, while the best network log-ARCH model ($k=3$-nearest neighbours with Euclidean distance, i.e. model B.3.1) reaches an average RMSFE of 2.44, which is about 15\% lower. The worst network model ($k=3$-nearest neighbours with correlation-based distance, i.e. model B.3.2), instead, has an average RMSFE of 2.55 thus providing a not negligible improvement in the forecasting accuracy compared to the log-ARCH. Considering MAFE loss, the log-ARCH has an average value of 2, which is much larger than 1.85, which is the average MAFE obtained with the best network model ($k=10$-nearest neighbours with volatility-based distance, i.e. model B.10.3). The worse network model in terms of MAFE ($k=3$-nearest neighbours with correlation-based distance, i.e. model B.3.2) has an average value of 1.94 which is still lower than the log-ARCH. 

From a first view, we can improve the forecasting accuracy of the log-ARCH by using any network structure. However, not all network structures are the same regarding out-of-sample forecasting accuracy. First, a comparison across network models highlights that the $k$-nearest neighbour approach leads to the construction of more effective financial networks from a financial point of view. In other words, fully connected networks obtained with the inverse distance approach are not the best forecasting choice. To be precise, the fully connected network constructed with the volatility-based approach \eqref{dist3} (model A.3) provides the best forecasts compared to the other two inverse-distance approaches (models A.1 and A.2) in terms of both RMSFE and MAFE. However, almost all $k$-nearest neighbours networks provide more accurate forecasts in the validation set.

\begin{sidewaystable*}
\caption{Comparison of the forecasting results using different edge weight matrices. The best RMSFE/MAFE in each row is printed in bold. We report the average performance across stocks and the best/worst cases (i.e., stocks) of the benchmark approach and the network ARCH.}\label{tab:weightcomparison}
\centering
\scalebox{0.9}{\begin{tabular}{l c ccc ccc ccc ccc}
\hline
                               &  log-ARCH    & \multicolumn{3}{c}{Inverse-distance weights} & \multicolumn{9}{c}{$k$-nearest-neighbours weights} \\
        Model                  &  Benchmark   &  A.1       &  A.2        &  A.3              & \multicolumn{3}{c}{B.$k$.1} & \multicolumn{3}{c}{B.$k$.2} & \multicolumn{3}{c}{B.$k$.3} \\
        $k$                    &              &            &             &                     & 3    & 5    & 10       & 3    & 5    & 10       & 3    & 5    & 10    \\
\hline
      Average RMSFE    & 2.8202 & 2.5324 & 2.5229 & 2.5194 & 2.4457 & 2.4904 & 2.4990 & 2.5476 & 2.4668 & 2.5017 & 2.4940 & 2.4571 & 2.4568 \\
   Average MAFE    &  2.0018 & 1.9000 & 1.8898 & 1.8896 & 1.8669 & 1.8905 & 1.8960 & 1.9369 & 1.8733 & 1.8920 & 1.8953 & 1.8675 & 1.8559 \\  [.2cm]
   RMSFE (best case benchmark, TRV)  & 2.3865 & 2.2025 & 2.1806 & 2.1683 & 2.1013 & 2.1772 & 2.1994 & 2.2076 & 2.1611 & 2.1780 & 2.1553 & 2.1151 & 2.1137 \\
   MAFE (best case benchmark, TRV)   & 1.7453 & 1.6967 & 1.6827 & 1.6809 & 1.6806 & 1.7186 & 1.7463 & 1.7317 & 1.7121 & 1.7224 & 1.6959 & 1.6829 & 1.6712 \\ [.2cm]
   RMSFE (worst case benchmark, MRK) & 3.6791 & 3.3520 & 3.3245 & 3.3166 & 3.2114 & 3.2920 & 3.2624 & 3.3363 & 3.2312 & 3.2561 & 3.2615 & 3.2467 & 3.2034 \\
   MAFE (worst case benchmark, MRK)  & 2.3385 & 2.1736 & 2.1502 & 2.1417 & 2.1052 & 2.1654 & 2.1257 & 2.2097 & 2.1004 & 2.1037 & 2.1422 & 2.1197 & 2.0519 \\  [.2cm]
   RMSFE (best case network, TRV)    & 2.3865 & 2.2025 & 2.1806 & 2.1683 & 2.1013 & 2.1772 & 2.1994 & 2.2076 & 2.1611 & 2.1780 & 2.1553 & 2.1151 & 2.1137 \\  
   MAFE (best case network, IBM)  & 1.9812 & 1.6766 & 1.6678 & 1.6709 & 1.6577 & 1.6865 & 1.6704 & 1.7009 & 1.6935 & 1.6697 & 1.6679 & 1.6761 & 1.6405 \\ [.2cm]
   RMSFE (worst case network, VZ)   & 3.6791 & 3.3520 & 3.3245 & 3.3166 & 3.2114 & 3.2920 & 3.2624 & 3.3363 & 3.2312 & 3.2561 & 3.2615 & 3.2467 & 3.2034 \\ 
   MAFE (worst case network, VZ)  & 2.3053 & 2.7820 & 2.6336 & 2.6580 & 2.3424 & 2.4007 & 2.5270 & 2.7314 & 2.2189 & 2.6203 & 2.4736 & 2.4006 & 2.3668 \\
\hline
\end{tabular}}
\end{sidewaystable*}

The superiority of the network-based log-ARCH models is also supported by predictive accuracy tests. In Tab. \ref{tab:dmtest1} and Tab. \ref{tab:dmtest2}, the results of the Diebold and Mariano \cite{diebold2002comparing} test are show for the comparison of the log-ARCH with the best and worse network models, respectively. Under the null hypothesis of the test we have that the benchmark performs equally or better than the network model. The predictive accuracy test are provided for all the $n=29$ stocks included in the sample and consider both squared and absolute errors. The larger the statistic, the larger is the improvement of the network model on the benchmark. 

The p-values in both Tab. \ref{tab:dmtest1} and Tab. \ref{tab:dmtest2} suggest that even the worse network approach provides statistically more accurate forecasts in out-of-sample than the benchmark. Interestingly, also outlier stocks highlighted in some network structures, e.g., BA and CRM for both returns or AXP and BA for volatility distances, are better predicted considering network information. This result is not straightforward, because it is reasonable to assume that the more distant the adjacent nodes are, the less relevant would be their information in forecasting.

\begin{table}[h]
\caption{\cite{diebold2002comparing} predictive accuracy test: benchmark vs best network model. Under the null the benchmark model (log-ARCH) has better or equal predictive accuracy than the best network log-ARCH approach.}
\label{tab:dmtest1}
\centering
\begin{tabular}{rrrrr}
  \hline
  & \multicolumn{2}{c}{Squared errors} &  \multicolumn{2}{c}{Absolute errors}\\
  Stock & DM stat & p-value & DM stat & p-value \\
  \hline
AAPL & 5.47 & 0.00 & 45.39 & 0.00 \\ 
  AMGN & 8.49 & 0.00 & 42.37 & 0.00 \\ 
  AXP & 3.76 & 0.00 & 41.08 & 0.00 \\ 
  BA & 3.34 & 0.00 & 38.80 & 0.00 \\ 
  CAT & 6.39 & 0.00 & 47.71 & 0.00 \\ 
  CRM & 2.99 & 0.00 & 47.03 & 0.00 \\ 
  CSCO & 7.90 & 0.00 & 50.06 & 0.00 \\ 
  CVX & 5.47 & 0.00 & 41.18 & 0.00 \\ 
  DIS & 4.05 & 0.00 & 41.18 & 0.00 \\ 
  GS & 6.53 & 0.00 & 44.79 & 0.00 \\ 
  HD & 4.67 & 0.00 & 43.63 & 0.00 \\ 
  HON & 6.35 & 0.00 & 41.39 & 0.00 \\ 
  IBM & 7.86 & 0.00 & 45.51 & 0.00 \\ 
  INTC & 4.72 & 0.00 & 40.16 & 0.00 \\ 
  JNJ & 5.84 & 0.00 & 42.88 & 0.00 \\ 
  JPM & 4.95 & 0.00 & 47.30 & 0.00 \\ 
  KO & 5.32 & 0.00 & 46.59 & 0.00 \\ 
  MCD & 7.21 & 0.00 & 51.82 & 0.00 \\ 
  MMM & 7.57 & 0.00 & 39.89 & 0.00 \\ 
  MRK & 7.64 & 0.00 & 47.51 & 0.00 \\ 
  MSFT & 4.15 & 0.00 & 52.28 & 0.00 \\ 
  NKE & 2.76 & 0.00 & 43.75 & 0.00 \\ 
  PG & 6.29 & 0.00 & 41.48 & 0.00 \\ 
  TRV & 5.09 & 0.00 & 34.94 & 0.00 \\ 
  UNH & 10.40 & 0.00 & 45.27 & 0.00 \\ 
  V & 4.13 & 0.00 & 49.77 & 0.00 \\ 
  VZ & 0.61 & 0.27 & 23.34 & 0.00 \\ 
  WBA & 7.36 & 0.00 & 43.15 & 0.00 \\ 
  WMT & 5.88 & 0.00 & 51.70 & 0.00 \\ 
   \hline
\end{tabular}
\end{table}

\begin{table}[h]
\caption{\cite{diebold2002comparing} predictive accuracy test: benchmark vs worst network model. Under the null the benchmark model (log-ARCH) has better or equal predictive accuracy than the worst network log-ARCH approach.}
\label{tab:dmtest2}
\centering
\begin{tabular}{rrrrr}
  \hline
  & \multicolumn{2}{c}{Squared errors} &  \multicolumn{2}{c}{Absolute errors}\\
  Stock & DM stat & p-value & DM stat & p-value \\
  \hline
AAPL & 5.12 & 0.00 & 45.39 & 0.00 \\ 
  AMGN & 4.62 & 0.00 & 42.37 & 0.00 \\ 
  AXP & 2.74 & 0.00 & 41.08 & 0.00 \\ 
  BA & 1.60 & 0.05 & 38.80 & 0.00 \\ 
  CAT & 5.77 & 0.00 & 47.71 & 0.00 \\ 
  CRM & 4.22 & 0.00 & 47.03 & 0.00 \\ 
  CSCO & 7.67 & 0.00 & 50.06 & 0.00 \\ 
  CVX & 3.46 & 0.00 & 41.18 & 0.00 \\ 
  DIS & 3.72 & 0.00 & 41.18 & 0.00 \\ 
  GS & 5.34 & 0.00 & 44.79 & 0.00 \\ 
  HD & 3.75 & 0.00 & 43.63 & 0.00 \\ 
  HON & 5.21 & 0.00 & 41.39 & 0.00 \\ 
  IBM & 6.94 & 0.00 & 45.51 & 0.00 \\ 
  INTC & 4.20 & 0.00 & 40.16 & 0.00 \\ 
  JNJ & 5.65 & 0.00 & 42.88 & 0.00 \\ 
  JPM & 4.16 & 0.00 & 47.30 & 0.00 \\ 
  KO & 5.13 & 0.00 & 46.59 & 0.00 \\ 
  MCD & 6.93 & 0.00 & 51.82 & 0.00 \\ 
  MMM & 6.92 & 0.00 & 39.89 & 0.00 \\ 
  MRK & 5.01 & 0.00 & 47.51 & 0.00 \\ 
  MSFT & 3.77 & 0.00 & 52.28 & 0.00 \\ 
  NKE & 2.72 & 0.00 & 43.75 & 0.00 \\ 
  PG & 5.20 & 0.00 & 41.48 & 0.00 \\ 
  TRV & 3.30 & 0.00 & 34.94 & 0.00 \\ 
  UNH & 9.66 & 0.00 & 45.27 & 0.00 \\ 
  V & 3.62 & 0.00 & 49.77 & 0.00 \\ 
  VZ & -3.52 & 0.00 & 23.34 & 0.00 \\ 
  WBA & 7.28 & 0.00 & 43.15 & 0.00 \\ 
  WMT & 5.24 & 0.00 & 51.70 & 0.00 \\ 
   \hline
\end{tabular}
\end{table}

\subsection{Does the network structure matter?}\label{sec:4.3}

The previous results show that network-based log-ARCH models are useful for predicting volatilities. However, Tab. \ref{tab:weightcomparison} highlights differences across the alternative network models in forecasting accuracy. For example, it is clear that the $k$-nearest neighbours network provide more accurate forecasts on average than inverse distance approaches. Thus, we may raise the question if the network structure matters? In other words, how the best-fitting network can be interpreted from a financial perspective.

To get more insights about the issue of finding the best network log-ARCH model, we apply the Model Confidence Set (MCS) procedure \cite{hansen2011model}, which aims at finding a smaller set of network models with statistically the same performances. The MCS procedures do not include network ARCH models with statistically lower forecasting performance in the superior sets. As in previous assessment, we consider the results of the MCS procedure in terms of both squared and absolute forecasting errors. The results are shown in Tab. \ref{tab:MCS1}, where Panel A reports the results under squared error loss, while Panel B under absolute error loss.

Tab. \ref{tab:MCS1} interestingly highlights that the superior set composition is the same regardless the adopted loss. Indeed, only three network-based ARCH models belong to the superior set and all of them are based on networks constructed according to the $k$-nearest neighbours procedure \eqref{eq:knn}. In particular, the models included in the superior set are the Euclidean distance, which is based on returns, with $k=3$, and the log-ARCH distance, which based on volatilities, with $k=5$ and $k=10$. 

\begin{table}[h]
\caption{Model Confidence Set: superior set of models - MSE and MAFE losses for the average errors. $e_{R,M}$ is the elimination rule, p-value is the MCS p-vlaue,  while Loss is the associated (MSE or MAFE) loss value. Rank provides the ranking of the models within the superior set in terms of the selected loss function.}
	\label{tab:MCS1}
	\centering
	\scalebox{0.9}{\begin{tabular}{lrrrr}
		\hline
		Network structure	& Rank & $e_{R,M}$ & p-val & Loss\\
		\hline
	\multicolumn{5}{l}{Panel A: squared errors' loss} \\
$k$-NN with \eqref{dist1} and $k=3$ (B.3.1)  & 1 & -1.75 & 1.00 & 6.033726\\ 
$k$-NN with \eqref{dist3} and $k=3$  (B.5.3)  & 3 & 1.11  & 0.39 & 6.093567\\ 
$k$-NN with \eqref{dist3} and $k=10$ (B.10.3)& 2 & 0.79 & 0.62 & 6.088585\\ 
		\hline
			\multicolumn{5}{l}{Panel B: absolute errors' loss} \\
$k$-NN with \eqref{dist1} and $k=3$ (B.3.1) & 2 & 0.85 & 0.56 & 1.866924\\ 
$k$-NN with \eqref{dist3} and $k=5$ (B.5.3) & 3 & 1.12 & 0.39 & 1.867474 \\ 
$k$-NN with \eqref{dist3} and $k=10$ (B.10.3) & 1 & -1.79 & 1.00 & 1.855948\\ 
	 \hline
	\end{tabular}}
\end{table}

However, it is interesting that the best model in the superior set differs according to squared and absolute forecasting errors. In the case of squared forecasting errors, the Euclidean distance \eqref{dist1} with $k=3$nearest neighbour provides the lowest loss, while in the case of absolute error, the best model is the $k=10$-nearest neighbour under log-ARCH distance \eqref{dist3}. Therefore, we can conclude that we only need the information from a few adjacent stocks in forecasting under returns-based networks. In contrast, information from a higher number of nodes is required for volatility-based networks. 

Overall, the results confirm that fully connected networks provide less accurate forecasts in out-of-sample; thus, $k$-nearest neighbours approaches should be preferred. Moreover, the results suggest that, although correlation-based approaches are the most widely used in the construction of financial networks, the correlation-based network ARCH is not included in the superior set. This means that information included in most correlated stocks is not as valuable for out-of-sample exercises as it appeared previously.  

In summary, Tab. \ref{tab:MCS1} shows that the network structure matters in terms of out-of-sample forecasting accuracy. Therefore,  researchers have to carefully specify the kind of network underlying the network ARCH model, even though the forecasting performance is good when not choosing the best network. Using a suitable network structure, it is possible to enhance the forecasting ability of the model further.

\subsection{Can the prediction performance be increased by considering multiple network definitions?}

In the end, we ask if it is possible improving the forecasting accuracy of the volatility with network log-ARCH models. A suitable idea is to use a combination of forecasts from the alternative models considered in the paper. Forecasting combination, also known as ensemble forecasting, is a technique used to improve the accuracy of predictions by combining multiple forecasts. The basic idea is that by combining the predictions of different models, the strengths of each can be leveraged to produce more accurate forecasting. By combining forecasts from multiple models, indeed, forecasters can reduce the risk of relying on a single model. The use of combination methods is nowadays widespread not only in economics \citep[e.g. see][]{proietti2021nowcasting}, but also in other research areas such as sociology \citep{tollenaar2013method}, epidemiology \citep{deb2022ensemble} and meteorology \citep{di2010bayesian}. In the context of volatility forecasting, ensemble techniques are also commonly considered \citep{becker2008combination}.

Although there are many ways of combining forecasts, we consider the three most common approaches, i.e. simple average, minimum-variance combination and constrained OLS, COLS, \cite[for details, see][]{timmermann2006forecast}. For the simple average method, forecasts are obtained by averaging the predictions from the alternative models. Although straightforward, there is wide evidence supporting the superiority of simple averaging compared with optimal combination approaches (i.e. the so-called ``forecast combination puzzle''). In the case of the minimum-variance approach, combination weights are obtained by minimising the resulting forecasting error variance. Then, in the constrained OLS combination the weights are obtained as the parameters of a linear regression, with a constraint on the parameters such that they sum up to one. In particular, the actual values of the time series to be predicted are regressed on the set of the alternative forecasts.

Below, we combine for each stock the forecasts obtained with both log-ARCH and the network log-ARCH, under the three aforementioned ensemble approaches. Then, we evaluate if combination further enhances forecasting accuracy in the validation sample. The results of the combination, considered in terms of both RMSFE and MAFE losses, are shown in Tab. \ref{tab:combination}. The last row of Tab. \ref{tab:combination} shows the average RMSFE and MAFE of each combination approach.

The ensemble forecast results can be compared with those of Tab. \ref{tab:weightcomparison}. In terms of average RMSFE and MAFE, the best combination approach is represented by the COLS. For example, the best network approach of Tab. \ref{tab:weightcomparison} has an average RMSFE of 2.45. With the COLS combination, we reduce it to 2.35, which is about 5\% lower. In terms of absolute errors, the best network achieves an average MAFE of 1.86, while with the COLS combination, we reduce the loss to 1.76, which is about 6\% smaller. Therefore, the improvements in the forecasting accuracy with the combination are not negligible. 

Interestingly, the simple average is the worst combination scheme to adopt. Contrary to this evidence, however, the best benchmark model performs (a bit) worse than this relatively easy combination approach. This suggests that we can improve forecasting accuracy with a low effort or, more generally, proficiently handling uncertainty about what model to use in a straightforward manner. The minimum-variance combination also improves forecasting accuracy, even if lower than the COLS.

Let us consider the results in terms of best and worse network cases. In the best case, the network log-ARCH provides an RMSFE of 2.11, but with COLS combination, we reduce the loss to 1.99, while we reduce the MAFE from 1.67 obtained with the best network model to 1.56 with COLS combination. In the worse case, we achieve an RMSFE of 3.20 with network log-ARCH, while with COLS combination, we reduce the loss to 2.59. Finally, in the case of MAFE loss, the two approaches perform similarly in the worst cases.

Based on the results presented in Table \ref{tab:combination}, combining forecasts obtained from the benchmark model and several network log-ARCH models appears advantageous rather than relying solely on a single model. This approach is beneficial for addressing the uncertainty associated with selecting an appropriate network structure. As discussed in Section \ref{sec:4.3}, the forecasts generated from different network structures are statistically different. While selecting the most suitable network structure is crucial, doing so ex-ante can be challenging and complex. By combining forecasts obtained from different networks, we can enhance the accuracy of out-of-sample forecasts and alleviate concerns about selecting the appropriate network structure.

\begin{table}[ht]
\centering
\caption{Ensemble forecasting for each stock. Results are reported inn terms of both RMSFE and MAFE accuracy metrics. Min. Var. indicates minimum-variance ensemble, while COLS is the Constrained OLS approach.}
\label{tab:combination}
\scalebox{0.90}{\begin{tabular}{rrrrrrr}
  \hline
  & \multicolumn{2}{c}{Simple Average} & \multicolumn{2}{c}{Min. Var.} & \multicolumn{2}{c}{COLS} \\ 
Stock & RMSFE & MAFE & RMSFE & MAFE & RMSFE & MAFE \\ 
  \hline
AAPL & 2.3522 & 1.8440 & 2.3263 & 1.8136 & 2.3209 & 1.8129 \\ 
  AMGN & 2.4915 & 1.8271 & 2.2754 & 1.6764 & 2.2399 & 1.6625 \\ 
  AXP & 2.3537 & 1.8207 & 2.2533 & 1.7024 & 2.2465 & 1.7019 \\ 
  BA & 2.5508 & 1.9933 & 2.4392 & 1.8734 & 2.4171 & 1.8550 \\ 
  CAT & 2.4031 & 1.8318 & 2.3514 & 1.8150 & 2.3381 & 1.8056 \\ 
  CRM & 2.5308 & 1.9261 & 2.4094 & 1.8271 & 2.4021 & 1.8318 \\ 
  CSCO & 2.2689 & 1.7101 & 2.2476 & 1.7247 & 2.2419 & 1.7226 \\ 
  CVX & 2.3810 & 1.8424 & 2.2831 & 1.7204 & 2.2431 & 1.6966 \\ 
  DIS & 2.6537 & 1.9728 & 2.6170 & 1.8982 & 2.5961 & 1.8808 \\ 
  GS & 2.3094 & 1.7037 & 2.2547 & 1.6654 & 2.2500 & 1.6718 \\ 
  HD & 2.2341 & 1.7274 & 2.2090 & 1.6892 & 2.2041 & 1.6922 \\ 
  HON & 2.3814 & 1.8023 & 2.3249 & 1.7485 & 2.3156 & 1.7451 \\ 
  IBM & 2.2139 & 1.6232 & 2.1901 & 1.6131 & 2.1803 & 1.6070 \\ 
  INTC & 2.4565 & 1.8827 & 2.4006 & 1.8101 & 2.3908 & 1.8050 \\ 
  JNJ & 2.6467 & 1.7668 & 2.6123 & 1.7642 & 2.5844 & 1.7545 \\ 
  JPM & 2.3275 & 1.7554 & 2.2674 & 1.7184 & 2.2550 & 1.7090 \\ 
  KO & 2.3686 & 1.7281 & 2.3178 & 1.6926 & 2.3154 & 1.6940 \\ 
  MCD & 2.3091 & 1.7454 & 2.2816 & 1.7295 & 2.2680 & 1.7209 \\ 
  MMM & 2.5672 & 1.9107 & 2.5072 & 1.9084 & 2.4964 & 1.9110 \\ 
  MRK & 3.2397 & 2.0732 & 3.1407 & 2.0694 & 3.1319 & 2.0739 \\ 
  MSFT & 2.6221 & 1.9454 & 2.5923 & 1.8978 & 2.5889 & 1.8992 \\ 
  NKE & 2.2772 & 1.7594 & 2.2148 & 1.6787 & 2.1956 & 1.6806 \\ 
  PG & 2.3088 & 1.7356 & 2.2350 & 1.7131 & 2.2317 & 1.7125 \\ 
  TRV & 2.1022 & 1.6435 & 2.0094 & 1.5689 & 1.9966 & 1.5610 \\ 
  UNH & 2.4257 & 1.7304 & 2.3590 & 1.7597 & 2.3481 & 1.7681 \\ 
  V & 2.1860 & 1.6969 & 2.0752 & 1.6361 & 2.0660 & 1.6341 \\ 
  VZ & 3.1516 & 2.4379 & 2.5011 & 1.9102 & 2.4768 & 1.8846 \\ 
  WBA & 2.4308 & 1.8502 & 2.3965 & 1.8413 & 2.3841 & 1.8278 \\ 
  WMT & 2.4345 & 1.8394 & 2.3940 & 1.8205 & 2.3793 & 1.8184 \\[.2cm]
Average & 2.4475 & 1.8319 & 2.3616 & 1.7685 & 2.3484 & 1.7635 \\ 
   \hline
\end{tabular}}
\end{table}

\section{Conclusion}\label{sec:conclusion}

In this paper, we propose a novel approach for forecasting volatility, which extends the log-ARCH to incorporate the network structure of financial time series. The stock market is well represented by networks, where stocks are the nodes, and the edges reflect the degree of similarity across them. By including the network connectives in the statistical model, we explicitly introduce the effect of instantaneous spillovers from adjacent nodes reflecting the simultaneity of investors' trading decisions. The information from adjacent nodes of a financial network can be used for forecasting purposes. 

There are many different ways of constructing financial networks. The paper evaluates the performances of twelve alternative network log-ARCH configurations. Inspired by time series clustering literature, three alternative dissimilarity definitions are considered for constructing the networks, i.e. Euclidean distance across returns, correlation-based and volatility-based. In addition, networks are considered both fully connected, employing an inverse distance approach, and not fully connected, utilising $k$-nearest neighbours with $k=\{3,5,10\}$. Finally, we use the proposed modelling approach to forecast the out-of-sample volatility of the stocks in the Dow Jones Index. 

First, we find that log ARCH models' forecasting accuracy significantly increases when including network information. This means that the information on adjacent network nodes is helpful in forecasting volatility. Moreover, we also show that the network structure matters in terms of out-of-sample forecasting accuracy. In particular, we find that networks constructed with inverse distance seem less effective in forecasting than those based on $k$-nearest neighbours \eqref{eq:knn}. Thus, fully connected networks appear not to be the best forecasting choice. Moreover, we can find three alternative Network log-ARCH models belonging to the superior set as suggested in \cite{hansen2011model}. Interestingly, none of these models adopts a correlation-based network, although this is one of the most common choices for constructing financial networks. Therefore, we suggest that practitioners carefully specify the kind of network underlying the network ARCH model.

\end{document}